\documentclass[12pt,twoside]{article}
\usepackage{amssymb}

\usepackage{amsmath}
\usepackage{theorem}

\def\1{{1\mskip-10mu1}}

\def\bea{\begin{eqnarray*}}
\def\eea{\end{eqnarray*}}
\def\bean{\begin{eqnarray}}
\def\eean{\end{eqnarray}}

\newtheorem{theorem}{Theorem}

\newtheorem{definition}[theorem]{Definition}
\newtheorem{example}[theorem]{Example}

\input{tcilatex}

\begin{document}

\title{{\LARGE A DESCRIPTION OF QUANTUM CHAOS}}
\author{Kei Inoue$\dagger $, Andrzej Kossakowski$\ddagger $ and Masanori Ohya%
$\dagger $ \\
$\dagger $Department of Information Sciences,\\
Science University of Tokyo,\\
Noda City, Chiba 278-8510, JAPAN.\\
$\ddagger $Institute of Physics, \\
N. Copernicus University, \\
Grudziadzka 5, 87-100 Torun, POLAND}
\date{}
\maketitle

\begin{abstract}
A measure describing the chaos of a dynamics was introduced by two
complexities in information dynamics, and it is called the chaos degree. In
particular, the entropic chaos degree has been used to characterized several
dynamical maps such that logistis, Baker's, Tinckerbel's in classical or
quantum systems. In this paper, we give a new treatment of quantum chaos by
defining the entropic chaos degree for quantum transition dynamics, and we
prove that every non-chaotic quantum dynamics, e.g., dissipative dynamics,
has zero chaos degree. A quantum spin 1/2 system is studied by our chaos
degree, and it is shown that this degree well describes the chaotic behavior
of the spin system.
\end{abstract}

\section{Introduction}

There exist several approaches in the study of chaotic behavior of dynamical
systems using the concepts such as (1) entropy and dynamical entropy, (2)
Chaitin's complexity, (3) Lyapunov exponent (4) fractal dimension (5)
bifurcation (6) ergodicity \cite{Aka,Ali,Dev,ENTS,Has,Maj} . But these
concepts are rather independently used in each field. In 1991, one of the
authors proposed Information Dynamics (ID for short) \cite{O2,O4,IKO} to try
to treat such chaotic behavior of systems from a common standing point. Then
a chaos degree to measure the chaos in dynamical systems is defined by means
of two complexities in ID\cite{O3,O4}. In particular, among several chaos
degrees, the entropic chaos degree was introduced in \cite{O5} and it is
applied to some dynamical systems\cite{O5,IOS,IOV1}. Recently, semiclassical
properties and chaos degree for quantum Baker's map has been considered in %
\cite{IOV1,IOV2}.

In this paper, we give a new treatment of quantum chaos by introducing the
chaos degree for quantum transition dynamics, and we prove some fundamental
properties for non-chaotic maps. Moreover we show, as an example, that our
chaos degree well describes chaotic behavior of spin systems.

\section{Entropic chaos degree}

In order to contain more general dynamics such as one in continuous systems,
we define the entropic chaos degree in C*-algebraic terminology. This
setting will be too general in the sequel discussions, but for mathematical
completeness including both classical and quantum systems, we start from the
C*-algebraic setting.

Let $(\mathcal{A},\mathfrak{S})$ be an input C* system and $(\overline{%
\mathcal{A}},\overline{{\mathfrak{S}}})$ be an output C* system; namely, $%
\mathcal{A}$ is a C* algebra with unit $I$ and $\mathfrak{S}$ is the set of
all states on $\mathcal{A}$. We assume $\overline{\mathcal{A}}=\mathcal{A}$
in the sequel for simplicity. For a weak* compact convex subset $\mathcal{S}$
(called the reference space) of $\mathfrak{S}$, take a state $\varphi $ from
the set $\mathcal{S}$ and let

\begin{equation*}
\varphi =\int_{\mathcal{S}}\omega d\mu _{\varphi }
\end{equation*}
be an extremal orthogonal decomposition of $\varphi $ in$\mathcal{\ S}$,
whose measure $\mu _{\varphi }$ describes a certain degree of mixture of $%
\varphi $ in the reference space $\mathcal{S}$. The measure $\mu _{\varphi }$
is not uniquely determined unless $\mathcal{S}$ is the Choquet simplex, so
that the set of all such measures is denoted by $M_{\varphi }\left( \mathcal{%
S}\right) .$

\begin{definition}
The entropic chaos degree with respect to $\varphi \in \mathcal{S}$ and a
channel $\Lambda ^{*},$ a map from $\frak{S}$ to $\overline{\frak{S}}$, is
defined by
\end{definition}

\begin{equation}
D^{\mathcal{S}}\left( \varphi ;\Lambda ^{*}\right) \equiv \inf \left\{ \int_{%
\mathcal{S}}S^{\mathcal{S}}\left( \Lambda ^{*}\varphi \right) d\mu _{\varphi
};\mu _{\varphi }\in M_{\varphi }\left( \mathcal{S}\right) \right\} \text{ }
\end{equation}
where $S^{\mathcal{S}}\left( \Lambda ^{*}\varphi \right) $ is the mixing
entropy of a state $\Lambda ^{*}\varphi $ in the reference space $\mathcal{S}
$ \cite{IKO}, hence it becomes von Neumann entropy when $\mathfrak{S}$ is
the set of all density operaotrs, or it does Shannon entropy when $\mathfrak{%
S}$ is the set of all probability distributions. This $D^{\mathcal{S}}\left(
\varphi ;\Lambda ^{*}\right) $ contains the classical chaos degree and the
quantum one. Now in the case of $\mathcal{S=}\mathfrak{S,}$ we simply denote 
$D^{\mathcal{S}}\left( \varphi ;\Lambda ^{*}\right) $ by $D\left( \varphi
;\Lambda ^{*}\right) .$

We use this degree to judge whether the dynamics $\Lambda ^{*}$ causes a
chaos or not as follows:

\begin{definition}
For a give state $\varphi ,$ a dynamics $\Lambda ^{*}$ causes chaos iff $D>0$
, and it does cause chaos (i.e., may be called stable) iff $D=0$ .
\end{definition}

In usual quantum system including classical discrete system, $\mathcal{A}$
is the set $\mathbf{B}\left( \mathcal{H}\right) $ of all bounded operators
on a Hilbert space $\mathcal{H}$ and $\mathfrak{S}$ is the set of all
density operators, in which an extremal decomposition of $\rho \in \mathfrak{%
S}$ is a Schatten decomposition $\rho =\sum_{k}p_{k}E_{k}$ (i.e., $\left\{
E_{k}\right\} $ are one dimensional orthogonal projections with $\sum
E_{k}=I),$ so that the entropic chaos degree is written as

\begin{equation}
D\left( \rho ;\Lambda ^{*}\right) \equiv \inf \left\{ \sum_{k}p_{k}S(\Lambda
^{*}E_{k});\left\{ E_{k}\right\} \right\} ,
\end{equation}
where the infimum is taken over all possible Schatten decompositions and $S$
is von Neumann entropy. Note that in classical discrete case, the Schatten
decomposition is unique $\rho =\sum_{k}p_{k}\delta _{k}$ with the delta
measure $\delta _{k}\left( j\right) \equiv \left\{ 
\begin{array}{ll}
1 & \left( k=j\right) \\ 
0 & \left( k\neq j\right)%
\end{array}
\right. ,$ and the entropic chaos degree is written by

\begin{equation}
D\left( \varphi ;\Lambda ^{*}\right) =\sum_{k}p_{k}S(\Lambda ^{*}\delta
_{k}),
\end{equation}
where $\rho $ is the probability distribution of the orbit obtained from a
dynamics of a system, and that dynamics generates the channel $\Lambda ^{*},$
whose details are discussed in Section 2. Before closing this section we
remark that in the case when a certain decomposition of the state $\rho $ is
fixed, say $\rho =\sum_{k}p_{k}\rho _{k}$, $\rho _{k}\in \mathfrak{S,}$ the
entropic chaos degree (ECD in the sequel) becomes

\begin{equation}
D\left( \varphi ;\Lambda ^{*}\right) =\sum_{k}p_{k}S(\Lambda ^{*}\rho _{k})
\end{equation}
without the infimum.

\section{Entropic chaos degree for classical dynamics}

Let us consider a map $f$ on $I\equiv \left[ a,b\right] ^{\mathbf{N}}\subset 
\mathbf{R}^{\mathbf{N}}$ with $x_{n+1}=f\left( x_{n}\right) $ (a difference
equation), $x_{0}\equiv x$. Take a finite partition of $I;$ $I\equiv
\bigcup_{k}B_{k}$ with $B_{i}\cap B_{j}=\emptyset $ $\left( i\neq j\right) . 
$ The state $\rho ^{\left( n\right) }$ at time $n$ determined by the
difference equation is the probability distribution $p^{\left( n\right) }$ $%
\equiv \left( p_{i}^{\left( n\right) }\right) $ of the orbit $\left\{
f^{n}\left( x\right) ;n=0,1,\cdots \right\} $, that is,

\begin{equation}
p_{i,B}^{\left( n\right) }\equiv \frac{1}{m+1}\sum_{k=n}^{m+n}1_{B_{i}}%
\left( f^{k}x\right) ,
\end{equation}
where $1_{A}$ is the characteristic function and $B\equiv \left\{
B_{i}\right\} $. When the initial value $x$ is distributed due to a measure $%
\nu $ on $I,$ the above $p_{i}^{\left( n\right) }$ is given as

\begin{equation}
p_{i,B}^{\left( n\right) }\equiv \frac{1}{m+1}\int_{I}%
\sum_{k=n}^{m+n}1_{B_{i}}\left( f^{k}x\right) d\nu .\text{ }
\end{equation}
The joint probability distribution $\left( p_{ij}^{\left( n,n+1\right)
}\right) $ between the time $n$ and $n+1$ is defined by

\begin{equation}
p_{ij,B}^{\left( n,n+1\right) }\equiv \frac{1}{m+1}\sum_{k=n}^{m+n}1_{B_{i}}%
\left( f^{k}x\right) 1_{B_{j}}\left( f^{k+1}x\right) \text{ }
\end{equation}
or

\begin{equation}
p_{ij,B}^{\left( n,n+1\right) }\equiv \frac{1}{m+1}\int_{I}%
\sum_{k=n}^{m+n}1_{B_{i}}\left( f^{k}x\right) 1_{B_{j}}\left(
f^{k+1}x\right) d\nu .\text{ }
\end{equation}
Then the channel $\Lambda _{n}^{*}$ at $n$ is defined by

\begin{equation}
\Lambda _{n,B}^{*}\equiv \left( \frac{p_{ij,B}^{\left( n,n+1\right) }}{%
p_{i,B}^{\left( n\right) }}\right) \Longrightarrow p_{B}^{\left( n+1\right)
}=\Lambda _{n,B}^{*}p_{B}^{\left( n\right) },
\end{equation}
and the chaos degree at time $n$ is given by

\begin{equation}
D\left( x;f\right) \equiv \sup_{\left\{ B_{i}\right\} }D\left( p_{B}^{\left(
n\right) };\Lambda _{n,B}^{*}\right) =\sup_{\left\{ B_{i}\right\}
}\sum_{i}p_{i,B}^{\left( n\right) }S(\Lambda _{n,B}^{*}\delta
_{i})=\sum_{i,j}p_{ij,B}^{\left( n,n+1\right) }\log \frac{p_{i,B}^{\left(
n\right) }}{p_{ij,B}^{\left( n,n+1\right) }}.\text{ }
\end{equation}
Therefore once we find a suitable partition $B$ such that $D\left(
p_{B}^{\left( n\right) };\Lambda _{n,B}^{*}\right) $ becomes positive, we
conclude that the dynamics $f$ produces chaos.

This entropic chaos degree has been applied to several dynamical maps such
logistic map, Baker's transformation and Tinkerbel map, and it could explain
their chaotic characters\cite{O4,IOS}. This chaos degree has several merits
to usual measures such as Lyapunov exponent.

\section{Entropic chaos degree for quantum dynamics}

Let us consider von Neumann-Liouville equation

\begin{equation}
i\frac{d\rho _{t}}{dt}=\left[ H\left( t\right) ,\rho _{t}\right]  \label{3.1}
\end{equation}

\noindent with the initial condition

\begin{equation}
\rho _{s}=\rho  \label{3.2}
\end{equation}

\noindent The solution of (\ref{3.1}) is given in the form

\begin{equation}
\rho _{t,s}=\Lambda _{t,s}^{*}\rho =U_{t,s}\rho U_{t,s}^{*}  \label{3.3}
\end{equation}
where

\begin{equation}
U_{t,s}=T\exp \left( -i\int_{s}^{t}H\left( t^{^{\prime }}\right)
dt^{^{\prime }}\right)  \label{3.4}
\end{equation}
and it follows from (\ref{3.4}) and (\ref{3.3}) that the relations

\begin{equation}
U_{t,s}U_{s,u}=U_{t,u},\qquad U_{t,t}=I,\qquad t\geq s\geq u  \label{3.5}
\end{equation}
and

\begin{equation}
\Lambda _{t,s}^{*}\Lambda _{s,u}^{*}=\Lambda _{t,u}^{*},\qquad \Lambda
_{t,t}^{*}=id,\qquad t\geq s\geq u  \label{3.6}
\end{equation}
hold.

From (\ref{3.5}) and (\ref{3.6}) one finds that

\begin{equation}
U_{t+\Delta ,s}=U_{t+\Delta ,t}U_{t,s}  \label{3.7}
\end{equation}
and

\begin{equation}
\Lambda _{t+\Delta ,s}^{*}=\Lambda _{t+\Delta ,t}^{*}\Lambda _{t,s}^{*}.
\label{3.8}
\end{equation}
That is, the relation between $U_{t+\Delta ,s}\left( \Lambda _{t+\Delta
,s}^{*}\right) $ and $U_{t,s}\left( \Lambda _{t,s}^{*}\right) $ is linear
one.

Let us put

\begin{eqnarray}
s &=&0,\quad t=n\tau  \notag \\
U_{n} &=&U_{n\tau ,0},\quad \Lambda _{n}^{*}=\Lambda _{n\tau ,0}^{*}
\label{3.9} \\
V_{n} &=&T\exp \left( -i\int_{\left( n-1\right) \tau }^{n\tau }H\left(
t^{^{\prime }}\right) dt^{^{\prime }}\right) .  \label{3.10}
\end{eqnarray}
Then one finds

\begin{eqnarray}
U_{n} &=&V_{n}V_{n-1}\cdots \cdots V_{1}  \label{3.11} \\
\Lambda _{n}^{*}\rho &=&V_{n}V_{n-1}\cdots \cdots V_{1}\rho \left(
V_{n}V_{n-1}\cdots \cdots V_{1}\right) ^{*}=U_{n}\rho U_{n}^{*}  \label{3.12}
\end{eqnarray}
The time dependence of $H\left( t\right) $ is generally very complicated.

One can consider as an example the following case. Let $H_{1},\ldots ,H_{n}$
be selfadjoint operators such that $I,F_{1},\ldots ,F_{n}$ are linearly
independent. Suppose that 
\begin{equation*}
H\left( t\right) =\sum_{k=1}^{N}c_{k}\left( t\right) F_{k}+H_{0}\qquad
for\quad t\geq 0
\end{equation*}
where $c_{1}\left( t\right) ,\ldots ,c_{n}\left( t\right) $ are solutions of
the equations

\begin{equation}
\frac{dc_{k}\left( t\right) }{dt}=f_{k}\left( c_{1}\left( t\right) ,\ldots
,c_{N}\left( t\right) \right) \qquad k=1,\ldots ,N  \label{3.13}
\end{equation}
with initial conditions

\begin{equation*}
c_{k}\left( 0\right) =c_{k}^{0}
\end{equation*}
and it is assumed that

\begin{equation*}
H\left( t\right) =H_{0}\qquad for\quad t<0.
\end{equation*}
The equations (\ref{3.13}) can lead to chaos. However its discrete version
is quite complicated.

Another example can be given in the form

\begin{equation}
H\left( t\right) =H_{0}+H_{n}\qquad for\quad \left( n-1\right) \tau \leq
t<n\tau ,\quad n=1,2,\ldots  \label{3.14}
\end{equation}
\begin{equation*}
H\left( t\right) =H_{0}\qquad for\quad t<0
\end{equation*}
and suppose that $H_{n}$ are determined as follows

\begin{equation}
H_{n+1}=G\left( H_{n}\right)  \label{3.15}
\end{equation}
or more explicitly

\begin{equation*}
H_{n}=\sum_{k=1}^{N}c_{n}^{\left( k\right) }\left( t\right) F_{k}
\end{equation*}
where

\begin{equation*}
c_{n+1}^{\left( k\right) }\left( t\right) =g_{k}\left( c_{n}^{\left(
1\right) },\ldots ,c_{n}^{\left( N\right) }\right) ,\quad n=1,2,\ldots
\end{equation*}
and $F_{k}$ are as above.

Thus the channel describing a discrete dynamics of a quantum systems as in
the above examples is written as 
\begin{equation}
\Lambda _{n+1}^{*}=\Theta _{n}^{*}\Lambda _{n}^{*}.  \label{3.16}
\end{equation}
and 
\begin{equation*}
\rho _{n+1}=\Lambda _{n+!}^{*}\rho =\Theta _{n}^{*}\rho _{n}.
\end{equation*}

It follows from the above examples that operators $V_{n}$ and consequently
maps $\Theta _{n}^{*}$ may inherit some chaotic properties of the equations (%
\ref{3.13}) or (\ref{3.15}). However it seems that the only way to
investigate the properties of $\Theta _{n}^{*}$ is to calculate expectation
values of some observables. The simple choice is to consider one observable.

Let $X\in B\left( \mathcal{H}\right) $ be an observable and $\rho \in 
\mathfrak{S}\left( \mathcal{H}\right) $ be an state of the system.

Let us consider the sequence

\begin{equation}
x_{n}=\mathrm{tr}X\Theta _{n}^{*}\rho ,  \label{3.17}
\end{equation}

\noindent where

\begin{equation}
\Theta _{n}^{*}\rho =V_{n}\rho V_{n}^{*}  \label{3.18}
\end{equation}

\noindent and

\begin{equation*}
V_{n}=T\exp \left( -i\int_{\left( n-1\right) \tau }^{n\tau }H\left(
t^{^{\prime }}\right) dt^{^{\prime }}\right) .
\end{equation*}

\noindent In the special case (\ref{3.14}) one has 
\begin{equation*}
V_{n}=e^{-i\tau \left( H_{0}+H_{n}\right) },\quad n=1,2,\ldots
\end{equation*}

The sequence $\left\{ x_{n}\right\} $ characterize the changes of $\Theta
_{n}^{*}$. Let $X$ and $\rho $ be fixed and take a proper $N$. Let $r,R$ be

\begin{equation}
r=\inf_{n}\mathrm{tr}X\Theta _{n}^{*}\rho =\inf_{n}x_{n},\qquad R=\sup_{n}%
\mathrm{tr}X\Theta _{n}^{*}\rho =\sup_{n}x_{n},\qquad n=1,2,\cdots ,N.
\label{3.19}
\end{equation}

\noindent In this case, the interval $\mathbf{I}$ in Section 2 can be
written by

\begin{equation}
\mathbf{I}=\left[ r,R\right] .  \label{3.20}
\end{equation}

Using the same way as we mentioned in Section 3, one can calculate the
entropic chaos degree $D$.

One can generalize the chaos degree taking the set $X_{1},X_{2},\ldots
,X_{L} $ $\left( L\in \mathbf{N}\right) $ of observables.

The quantum entropic chaos degree is applied to analyze quantum spin systems
and quantum baker's map, and we could measure the chaos of these systems.

\section{Properties of entropic chaos degree for quantum dynamics}

In this section we explain some properties of the entropic chaos degree for
quantum dynamics. We have the following theorem.

\begin{theorem}
For any $\rho \in \frak{S}\left( \mathcal{H}\right) $, $\Lambda _{n}^{*}:$ $%
\frak{S}\left( \mathcal{H}\right) \rightarrow \frak{S}\left( \mathcal{H}%
\right) $, $n=1,2,\cdots $, we have:
\end{theorem}

\begin{itemize}
\item[(1)] Let $U$ be a unitary operator. If $\Lambda _{n}^{*}\rho
=U^{n}\rho U^{*n}$, then $D\left( \rho ;\Lambda _{n}^{*}\right) =0$,

\item[(2)] If $\Lambda _{n}^{*}\rho =\rho $, then $D\left( \rho ;\Lambda
_{n}^{*}\right) =0$,

\item[(3)] Let $\rho _{0}$ be a fixed state on $\mathcal{H}$. If $\Lambda
_{n}^{*}\rho =\rho _{0}$, then $D\left( \rho ;\Lambda _{n}^{*}\right) =0$,

\item[(4)] Let $\left\{ P_{n}\right\} $ be one dimensional projections such
that $\sum_{k}P_{k}=I$. If $\Lambda _{n}^{*}\rho =\sum_{k_{1\cdots
}k_{n}}P_{k_{n}}P_{k_{n-1}}\cdots P_{k_{1}}\rho P_{k_{1}}\cdots
P_{k_{n-1}}P_{k_{n}}$, then $D\left( \rho ;\Lambda _{n}^{*}\right) =0.$
\end{itemize}

\noindent \textbf{Proof: }Let $X$ be an observable, $r,R$ be real numbers
given by (\ref{3.19}), $n,m,$ $M$ be large natural numbers and

\begin{equation*}
B_{i}=\left[ \frac{i}{M}\left( R-r\right) +r,\frac{i+1}{M}\left( R-r\right)
+r\right)
\end{equation*}

\noindent for $i=0,\cdots ,M-2,$

\begin{equation*}
B_{M-1}=\left[ \frac{M-1}{M}\left( R-r\right) +r,R\right] .
\end{equation*}

\noindent (1) By a direct calculation, we obtain

\begin{equation*}
x_{k}=\mathrm{tr}X\Theta _{k}^{*}\rho =\mathrm{tr}XU\rho U^{*},\qquad
k=1,2,\cdots .
\end{equation*}

\noindent Let $B_{n_{0}}$be a domain including the point $\mathrm{tr}XU\rho
U^{*}$. Then we have

\begin{equation*}
x_{k}\in B_{n_{0}}
\end{equation*}
because $x_{k}$ is independent of $k$. One finds that

\begin{eqnarray}
p_{n_{0},B}^{\left( n\right) } &=&\frac{1}{m+1}\sum_{k=n}^{m+n}1_{B_{n_{0}}}%
\left( x_{k}\right) =1  \label{4.1} \\
p_{n_{0}n_{0},B}^{\left( n,n+1\right) } &\equiv &\frac{1}{m+1}%
\sum_{k=n}^{m+n}1_{B_{n_{0}}}\left( x_{k}\right) 1_{B_{n_{0}}}\left(
x_{k+1}\right)  \label{4.2}
\end{eqnarray}

\noindent (\ref{4.1}) and (\ref{4.2}) imply that

\begin{equation*}
D\left( p_{B}^{\left( n\right) };\Lambda _{n,B}^{*}\right) =0
\end{equation*}
for any partition $B,$ hence 
\begin{equation*}
D\left( \rho ;\Lambda _{n}^{*}\right) =0.
\end{equation*}

\noindent (2) Note that

\begin{equation*}
x_{k}=\mathrm{tr}X\Theta _{k}^{*}\rho =\mathrm{tr}A\rho ,\qquad k=1,2,\cdots
.
\end{equation*}

\noindent for $k\in \mathbf{N}$.

Let $B_{n_{1}}$ be a domain including the point $tr\left( X\rho \right) $.
Then we have

\begin{equation*}
x_{k}\in B_{n_{1}}
\end{equation*}
because $x_{k}$ is independent of $k$. One finds that

One can show that

\begin{equation}
p_{n_{1,}B}^{\left( n\right) }=p_{n_{1}n_{1},B}^{\left( n,n+1\right) }=1
\label{4.3}
\end{equation}

This equation (\ref{4.3}) implies

\begin{equation*}
D\left( \rho ;\Lambda _{n}^{*}\right) =0.
\end{equation*}

\noindent (3) A direct calculation yields 
\begin{equation*}
x_{k}=\mathrm{tr}X\Theta _{k}^{*}\rho =\mathrm{tr}X\rho _{0},\qquad
k=1,2,\cdots .
\end{equation*}

\noindent for $k\in \mathbf{N}$.

Let $B_{n_{2}}$ be a domain including the point $\mathrm{tr}A\rho _{0}$.
Then we have

\begin{equation*}
x_{k}\in B_{n_{2}}
\end{equation*}
because $x_{k}$ is independent of $k$. It follows that 
\begin{equation*}
p_{n_{2},B}^{\left( n\right) }=p_{n_{2}n_{2},B}^{\left( n,n+1\right) }=1
\end{equation*}

\noindent Similarly (2), we obtain

\begin{equation*}
D\left( \rho ;\Lambda _{n}^{*}\right) =0.
\end{equation*}

\noindent (4) By a direct calculation, we have

\begin{equation*}
x_{k}=\mathrm{tr}X\Theta _{k}^{*}\rho =\mathrm{tr}X\left( \sum_{j}P_{j}\rho
P_{j}\right) ,\qquad k=1,2,\cdots .
\end{equation*}

\noindent for $k\in \mathbf{N}$.

Let $B_{n_{3}}$ be a domain including the point $\mathrm{tr}X\left(
\sum_{j}P_{j}\rho P_{j}\right) $. Then we have

\begin{equation*}
x_{k}\in B_{n_{3}}
\end{equation*}
because $x_{k}$ is independent of $k$. One can show that

\begin{equation}
p_{n_{3},B}^{\left( n\right) }=p_{n_{3}n_{3},B}^{\left( n,n+1\right) }=1
\label{4.4}
\end{equation}

This equation (\ref{4.4}) implies

\begin{equation*}
D\left( \rho ;\Lambda _{n}^{*}\right) =0.\text{ }\blacksquare
\end{equation*}

\section{Entropic chaos degree for quantum dynamics in spin 1/2 system}

Let us study a spin 1/2 system in external magnetic field. The Hamiltonian
of the system has the form

\begin{equation}
H\left( t\right) =\frac{e\hbar }{2mc}\vec{\sigma}\vec{B}\left( t\right) =%
\frac{1}{2}\vec{\sigma}\vec{y}_{t}  \label{5.1}
\end{equation}
where

\begin{equation}
\vec{y}_{t}=\frac{e\hbar }{mc}\vec{B}\left( t\right) \in \mathbf{R}^{3}.
\label{5.2}
\end{equation}
Following the scheme presented in Section 4, one can choose operators $%
F_{1},F_{2},F_{3}$ in (\ref{3.13}) as $\sigma _{1},\sigma _{2},\sigma _{3},$
then the equations (\ref{3.13}) will take the form

\begin{equation}
\frac{dy_{t}^{\left( k\right) }}{dt}=Y_{k}\left( \vec{y}_{t}\right) ,\quad
k=1,2,3,\quad \vec{y}_{t_{0}}=\vec{y}_{0}  \label{5.3}
\end{equation}
and 
\begin{equation*}
\vec{y}_{t}=\vec{c}\qquad for\quad t<0.
\end{equation*}
In what follows we will proceed according to the second example (\ref{3.15})
one has 
\begin{equation}
\vec{y}_{n+1}=\vec{Z}\left( \vec{y}_{n}\right) ,\quad n=0,1,2,\ldots
\label{5.4}
\end{equation}
Denoting 
\begin{equation*}
\omega _{n}=\left| \vec{y}_{n}\right| ,\quad \vec{e}_{n}=\frac{\vec{y}_{n}}{%
\left| \vec{y}_{n}\right| },
\end{equation*}
one has 
\begin{equation*}
\vec{y}_{n}=\omega _{n}\vec{e}_{n}.
\end{equation*}
In this case 
\begin{equation*}
V_{n}=\exp \left( -i\frac{\omega _{n}\tau }{2}\left( \vec{e}_{n},\vec{\sigma}%
\right) \right) .
\end{equation*}
Instead of (\ref{5.4}) we can rewrite 
\begin{equation}
\vec{e}_{n+1}=\vec{e}\left[ \vec{e}_{n}\right] ,\quad \vec{e}_{n}=\left(
e_{n}^{\left( 1\right) },e_{n}^{\left( 2\right) },e_{n}^{\left( 3\right)
}\right) .  \label{5.5}
\end{equation}
Any observable can be written in the form $X=\vec{a}\vec{\sigma},\vec{a}\in 
\mathbf{R}^{3}$, then one finds 
\begin{equation*}
V_{n}^{*}\vec{a}\vec{\sigma}V_{n}=\Theta _{n}\vec{a}\vec{\sigma}=\vec{\sigma}%
R\left( \omega \tau ,\vec{e}_{n}\right) \vec{a}
\end{equation*}

\noindent where 
\begin{equation*}
R\left( \omega \tau ,\vec{e}_{n}\right) \vec{a}=\left[ \vec{a}-\vec{e}%
_{n}\left( \vec{e}_{n}\vec{a}\right) \right] \cos \omega \tau +\vec{e}%
_{n}\left( \vec{e}_{n}\vec{a}\right) -\left( \vec{e}_{n}\wedge \vec{a}%
\right) \sin \omega \tau .
\end{equation*}

\noindent Using (\ref{3.17}), (\ref{3.18}) ,(\ref{5.5}) and the standard
form of state in $\mathbf{C}^{2}$

\begin{equation}
\rho =\frac{1}{2}\left( I+\vec{\sigma}\vec{\rho}\right) ,\qquad \vec{\rho}%
\in \mathbf{R}^{3},\qquad \left\| \vec{\rho}\right\| \leq 1  \label{5.6}
\end{equation}
One finds that

\begin{eqnarray}
x_{n}\left( X,\rho \right) &=&\left( \vec{\rho},R\left( \omega \tau ,\vec{e}%
_{n}\right) \vec{a}\right) =x_{n}\left( \vec{a},\vec{\rho}\right)
\label{5.7} \\
&=&\vec{\rho}\left[ \vec{a}-\vec{e}_{n}\left( \vec{e}_{n}\vec{a}\right) %
\right] \cos \omega \tau +\vec{\rho}\vec{e}_{n}\left( \vec{e}_{n}\vec{a}%
\right) -\vec{\rho}\left( \vec{e}_{n}\wedge \vec{a}\right) \sin \omega \tau
\label{5.8}
\end{eqnarray}
The special examples of (\ref{5.5}) are the followings:

\begin{example}
\begin{eqnarray*}
e_{n+1}^{\left( 1\right) } &=&\left( 1-\cos \theta \right) e_{n}^{\left(
3\right) }e_{n}^{\left( 1\right) }-\left( \sin \theta \right) e_{n}^{\left(
2\right) } \\
e_{n+1}^{\left( 2\right) } &=&\left( 1-\cos \theta \right) e_{n}^{\left(
3\right) }e_{n}^{\left( 2\right) }+\left( \sin \theta \right) e_{n}^{\left(
1\right) } \\
e_{n+1}^{\left( 3\right) } &=&\cos \theta +\left( 1-\cos \theta \right)
\left( e_{n}^{\left( 3\right) }\right) ^{2}
\end{eqnarray*}

The last relation can be written in the form 
\[
z_{n+1}=4bz_{n}\left( 1-z_{n}\right) 
\]

where 
\[
z_{n}=\frac{1}{2}\left( 1-e_{n}^{\left( 3\right) }\right) ,b=\left( \sin 
\frac{\theta }{2}\right) ^{2}. 
\]
\end{example}

\begin{example}
\begin{eqnarray*}
e_{n+1}^{\left( 1\right) } &=&\left[ -1+2\left( 1-\cos \theta \right) \left(
e_{n}^{\left( 3\right) }\right) ^{2}\right] e_{n}^{\left( 1\right) }-2\left(
\sin \theta \right) e_{n}^{\left( 3\right) }e_{n}^{\left( 2\right) } \\
e_{n+1}^{\left( 2\right) } &=&\left[ -1+2\left( 1-\cos \theta \right) \left(
e_{n}^{\left( 3\right) }\right) ^{2}\right] e_{n}^{\left( 2\right) }-2\left(
\sin \theta \right) e_{n}^{\left( 3\right) }e_{n}^{\left( 1\right) } \\
e_{n+1}^{\left( 3\right) } &=&\left( 1-a\right) e_{n}^{\left( 3\right)
}+a\left( e_{n}^{\left( 3\right) }\right) ^{2}
\end{eqnarray*}

with

\[
a=2\left( 1-\cos \theta \right) ,\qquad 0\leq a\leq 4 
\]
\end{example}

Choosing $\vec{a}=\left( 0,0,1\right) =\vec{a}_{0}$ and $\vec{\rho}=\left(
0,0,1\right) =\vec{\rho}_{0},$ we have 
\begin{eqnarray*}
x_{n}\left( \vec{a}_{0},\vec{\rho}_{0}\right) &=&\left[ 1-\left(
e_{n}^{\left( 3\right) }\right) ^{2}\right] \cos \omega \tau +\left(
e_{n}^{\left( 3\right) }\right) ^{2} \\
&=&\cos \omega \tau +\left( e_{n}^{\left( 3\right) }\right) ^{2}\left(
1-\cos \omega \tau \right) .
\end{eqnarray*}
It is clear that \noindent in order to investigate the properties of the
sequence $x_{n}\left( \vec{a}_{0},\vec{\rho}_{0}\right) $, it is enough to
consider the sequence

\begin{equation*}
x_{n}^{^{\prime }}=\left( e_{n}^{\left( 3\right) }\right) ^{2}
\end{equation*}

\noindent which will be chaotic.

The entropic chaos degree $D$ for the above two examples are shown in Fig.
6.1 and 6.2 under initial value $e_{0}^{\left( i\right) }=1/\sqrt{3}\left(
i=1,2,3\right) $.

\FRAME{dtbpF}{5.1526in}{3.3745in}{0pt}{}{}{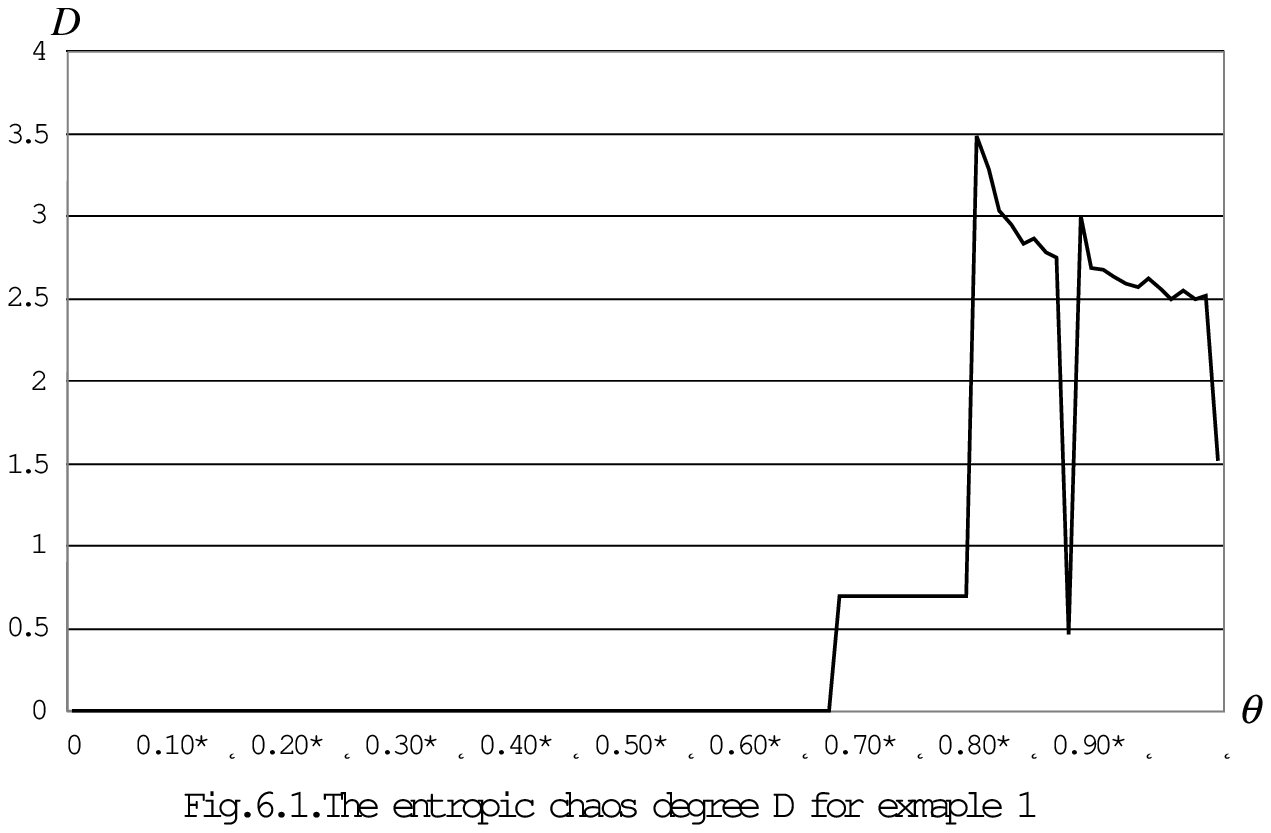}{\special{language
"Scientific Word";type "GRAPHIC";maintain-aspect-ratio TRUE;display
"USEDEF";valid_file "F";width 5.1526in;height 3.3745in;depth
0pt;original-width 5.0981in;original-height 3.3278in;cropleft "0";croptop
"1";cropright "1";cropbottom "0";filename 'fig1.eps';file-properties
"XNPEU";}}

\FRAME{dtbpF}{13.0787cm}{8.5427cm}{0pt}{}{}{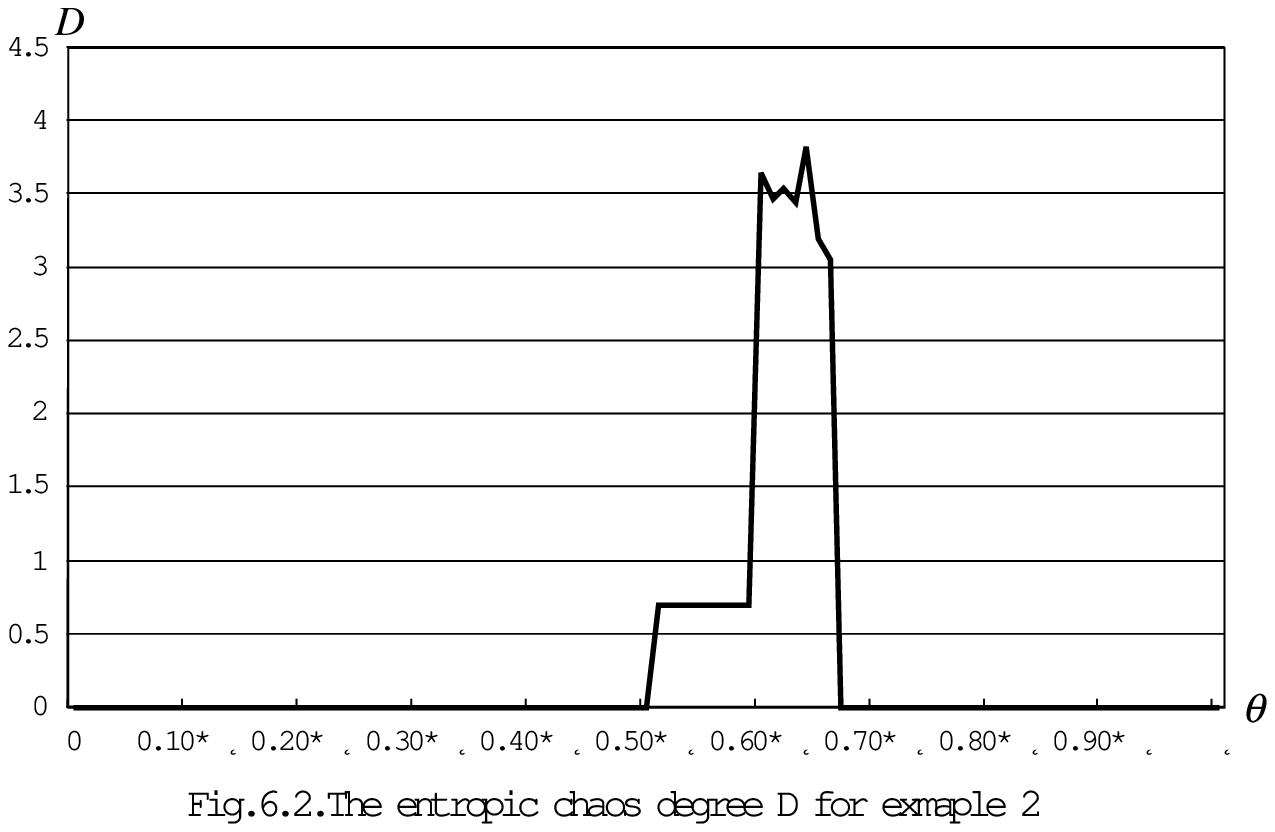}{\special{language
"Scientific Word";type "GRAPHIC";maintain-aspect-ratio TRUE;display
"USEDEF";valid_file "F";width 13.0787cm;height 8.5427cm;depth
0pt;original-width 5.0981in;original-height 3.3157in;cropleft "0";croptop
"1.0004";cropright "0.9993";cropbottom "0";filename
'fig2.eps';file-properties "XNPEU";}}

\end{document}